\documentclass[%
 aip,
 amsmath,amssymb,
 reprint,%
]{revtex4-1}

\usepackage{graphicx}
\usepackage{dcolumn}
\usepackage{bm}

\usepackage[utf8]{inputenc}
\usepackage[T1]{fontenc}
\usepackage{mathptmx}
\usepackage{etoolbox}
\usepackage{natbib}
\makeatletter
\def\@email#1#2{%
 \endgroup
 \patchcmd{\titleblock@produce}
  {\frontmatter@RRAPformat}
  {\frontmatter@RRAPformat{\produce@RRAP{*#1\href{mailto:#2}{#2}}}\frontmatter@RRAPformat}
  {}{}
}%
\makeatother
\begin{document}

\preprint{AIP/123-QED}

\title{The role of particle feedback on particle acceleration in magnetic reconnection}

\author{Shi Min Liang}
\textsuperscript{1}\affiliation{School of Mathematics and Computational Sciences, Xiangtan University, Xiangtan, Hunan 411105, People’s Republic of China}
\textsuperscript{2}\affiliation{Department of Physics, Xiangtan University, Xiangtan, Hunan 411105, People’s Republic of China}
\email{202331510127@smail.xtu.edu.cn}
\author{Nian Yu Yi}%
\textsuperscript{1}\affiliation{School of Mathematics and Computational Sciences, Xiangtan University, Xiangtan, Hunan 411105, People’s Republic of China}
\textsuperscript{3}\affiliation{Hunan Key Laboratory for Computation and Simulation in Science and Engineering, Xiangtan 411105, People’s Republic of China}
\textsuperscript{4}\affiliation{Hunan National Center for Applied Mathematics, Xiangtan 411105, People’s Republic of China}%

\date{\today}

\begin{abstract}

Magnetic reconnection is a ubiquitous process in astrophysical plasmas and an efficient mechanism for particle acceleration. Using 2.5D magnetohydrodynamic (MHD) simulations with a co-evolving fluid-particle framework, we investigate how particle feedback affects reconnection and acceleration. Our simulations demonstrate that particle feedback to the fluid amplifies shear flows within magnetic islands, which strengthens the convective electric field and thereby boosts particle acceleration. This mechanism results in a higher maximum particle energy and a harder nonthermal energy spectrum. The guide field suppresses both the increase in gas internal energy and particle acceleration. These findings highlight the complex interplay between feedback, guide fields, and reconnection dynamics.

\end{abstract}

\maketitle

\section{\label{sec:intro}Introduction}
The study of high-energy plasma dynamics is crucial for understanding the complex phenomena associated with high-energy astrophysical sources, such as pulsars,\cite{Uzdensky2014} supernova remnants,\cite{Cerutti2014a} and active galactic nuclei.\cite{Kadowaki2015} These environments are characterized by strong electromagnetic fields that dominate plasma behavior.  Observations of $\gamma$-ray sources emitting photons up to 1.4 PeV indicate the presence of cosmic ray (CR) factories,\cite{Cao2021} or PeVatrons, that accelerate particles to extreme energies. However, the precise mechanisms of particle acceleration in these environments remain poorly understood.

Magnetic reconnection is widely considered one of the most effective mechanisms for particle acceleration.\cite{deGouveiadalPino2005} Charged particles can be accelerated through several mechanisms: direct acceleration by the electric field in the current sheet,\cite{Sironi2014} first-order Fermi acceleration as particles converge into the reconnection region,\cite{deGouveiadalPino2005} and trapping and acceleration within contracting magnetic islands.\cite{Drake2010,Guo2014,Guo2015} These findings were mainly achieved through particle-in-cell (PIC) simulations and have been extended to MHD and test-particle simulations,\cite{Liu2009,Kowal2011,Ripperda2017a} the results of which are consistent with those of PIC simulations. For example, \cite{deGouveiaDalPino2013} proposed a model of 2D contracting magnetic islands, where particles trapped within the islands can be accelerated, gaining energy exponentially due to island mergers.

However, conventional numerical approaches, such as PIC methods, are fundamentally limited to kinetic scales, whereas astrophysical systems typically span vastly larger spatial dimensions. Although the MHD+test particle method is applicable to larger spatial scales than traditional PIC simulations, it faces significant challenges in resolving the ion skin depth, a critical scale that is typically several orders of magnitude smaller than the system size in astrophysical contexts.\cite{Bai2015,Mignone2018} Moreover, conventional MHD+test particle methods typically investigate particle acceleration using frozen turbulence data, where the fluid is treated as a static background structure. This approximation neglects the interactions between the particles and fluid dynamics.\cite{Liu2009,Kowal2011,Ripperda2017a} However, these interactions are essential for a complete description of energy transfer and particle acceleration mechanisms in astrophysical plasmas.\cite{Bai2015,Seo2024}

In this work, we adopt the MHD-PIC method for numerical simulations of magnetic reconnection. The MHD-PIC method combines the advantages of MHD and PIC techniques, enabling the simultaneous evolution of both fluid and particles while capturing their mutual interactions. We employ a 2.5D MHD-PIC approach, where particles are injected into the fluid, allowing the co-evolution of particles and fluid to investigate the role of particle feedback in reconnection-driven acceleration. Our approach allows for a more realistic representation of the mutual interactions between particles and the fluid, providing a significant advancement over traditional methods. The layout of this paper is as follows. We briefly explain the MHD-PIC method in Section \ref{Sect:MHD-PIC module} and present our simulation setup in Section \ref{Sect:numerical setting}. Section \ref{sect:Res} provides the results of the numerical simulation. Finally, we give the discussion and summary in Sections \ref{Sec:Discussion} and \ref{Sec:Summary}.

\section{\label{sec:simul} Simulation Method}
\subsection{MHD-PIC Module} \label{Sect:MHD-PIC module}
The MHD-PIC module implemented in the PLUTO code\cite{Mignone2018} provides a self-consistent framework for simulating magnetized plasma systems consisting of both thermal and nonthermal components. This hybrid approach solves the MHD equations for the thermal fluid while simultaneously tracking the dynamics of CR particles using PIC techniques. The module incorporates two-way coupling between the components through momentum-energy feedback and includes the CR-induced Hall term in Ohm's law.\cite{Bai2015}

The numerical implementation comprises three primary components: (1) a finite-volume solver for the MHD equations, (2) a particle integrator for the CR dynamics, and (3) a time-stepping scheme that maintains synchronization between the fluid and particle components. The MHD solver enforces the divergence-free condition through either constrained transport or divergence-cleaning methods. For particle integration, the module offers both time-reversible PIC algorithms and Runge-Kutta schemes. The particles represent phase-space clouds of physical particles, enabling efficient large-scale simulations while preserving kinetic effects. For details of this module, see references \onlinecite{Bai2015} and \onlinecite{Mignone2018}.

This approach is particularly suited for systems where the CR gyroradius (typically representing energetic protons) significantly exceeds the ion skin depth of the background plasma. It is important to note that while the MHD-PIC model self-consistently captures the momentum and energy feedback from CR particles to the background fluid, it does not resolve the pressure anisotropy of individual particles. This approximation is valid under the assumption of efficient pitch-angle scattering, which maintains near-isotropy in the particle distribution, but may omit certain kinetic instabilities driven by pressure anisotropy.
The module has been successfully validated against standard test problems, including Bell instability and shock acceleration scenarios, showing excellent agreement with the results of the $Athena++$ code.\cite{Bai2015} Computational performance analysis demonstrates efficient parallel scaling, with CPU utilization remaining stable above 0.8 in production runs.\cite{Mignone2018} 

\subsection{Numerical Setting}\label{Sect:numerical setting}

We consider a single conductive fluid, ignoring the dynamic effects and electronic physics of the fluid. We conducted numerical simulations using the MHD-PIC method based on the PLUTO code\cite{Mignone2007,Mignone2018} and solved the following dimensionless equations:
\begin{equation} \label{1}
		\frac{\partial \rho}{\partial t}+\nabla \cdot{(\rho \bm{v}_{\rm g})}=0,
\end{equation} 
\begin{equation} \label{2}
	\frac{\partial \bm{m}}{\partial t}+\nabla \cdot[\bm{m} \bm{v}_{\rm g} -\bm {BB}+\mathbf{I}(p+\frac{\bm{B}^2}{2})]=\bm{f}_{\rm p},
\end{equation}
\begin{equation} \label{3}
	\frac{\partial E_{\rm t}}{\partial t}+\nabla \cdot [(\frac{1}{2}\rho \bm{v}_{\rm g}^2+\rho \epsilon+p)\bm{v}_{\rm g} + c\bm{E} \times \bm{B}]=W_{\rm p},
\end{equation}
\begin{equation} \label{4}
	\frac{\partial \bm{B}}{\partial t}+\nabla \times(c\bm{E})=0,
\end{equation}
\begin{equation} \label{5}
	\nabla \cdot \bm{B}=0.
\end{equation}
These equations are the continuity, momentum, energy, induction, and solenoidal condition, respectively. Here, $\rho$ is the mass density, $\bm{m}=\rho \bm{v}_{\rm g}$ the momentum density, $I$ unit tensor, $\epsilon$ specific internal energy, $\bm{v}_{\rm g}$ the gas velocity, $p$ the gas pressure, $\bm{B}$ the magnetic field, and $E_{\rm t} = \rho \epsilon + \bm{m}^2/2 \rho+ \bm{B}^2/2$ the total energy density. The $\bm{f}_{\rm p}$ and $W_{\rm p}$ in Eqs. (\ref{2}) and (\ref{3}) are coupling terms, respectively, the force and work (per unit volume) performed by the particles onto the gas. The subscript $\rm p$ refers to a quantity related to particles.
The electric field $\bm E$ is specified by Ohm's law, whose form depends on the particular model at hand. 

The $\bm{f}_{\rm p}$ and $W_{\rm p}$ are given by:
\begin{equation}\label{6}
    \bm{f}_{\rm p} = -{\bm F}_{\rm p} = -q_{\rm p}\bm{E}-\frac{1}{c}\bm{J}_{\rm p}\times \bm B,
\end{equation}
\begin{equation}\label{7}
    W_{\rm p} = -\bm{v}_{\rm g}\cdot \bm{F}_{\rm p},
\end{equation}
where $q_{\rm p}$ and $\bm{J}_{\rm p}$ are the CR charge and current densities, respectively. The electric field directly comes from the equation of motion of massless electrons (after ignoring Hall and electron pressure terms):
\begin{equation}
    c\bm{E}=-\bm{v}_{\rm g} \times \bm{B}-R(\bm{v}_{\rm p}-\bm{v}_{\rm g})\times \bm B,
\end{equation}
where $c$ is the speed of light and a value can be manually specified. For consistency, this value must be greater than any characteristic velocity of the MHD. $R$ represents the ratio of CR charge density to total charge density:
\begin{equation}
    R = \frac{q_{\rm p}}{q_{\rm i}+q_{\rm p}}=\frac{\alpha_{\rm p}\varrho_{\rm p}}{\alpha_{\rm g}\rho_{\rm g}+\alpha_{\rm p}\varrho_{\rm p}},
\end{equation}
where $\alpha$ is the charge-to-mass ratio $\alpha_{\rm p}=(e/mc)_{\rm p}$ and $\alpha_{\rm g}=(e/mc)_{\rm g}$, $e$ is the elementary charge. 

Our numerical simulations are performed in a 2D domain with physical dimensions $1.0 L \times 0.5 L$, where the characteristic length scale $L$ is defined as $L = 10^4{V_{\rm A}/\omega_{\rm p}}$ with the light speed of $c=10^4V_{\rm A}$ and the plasma frequency of $\omega_{\rm p} = \sqrt{\rho q^2/m_{\rm p}}$ ($q$, $m_{\rm p}$ are the charge and mass of the proton, respectively). The configuration of the initial magnetic field is considered as Harris type\cite{Harris1962} by 
\begin{equation} \label{8}
	\bm B = B_0 {\rm tanh}\frac{y}{w}\bm e_x,
\end{equation}
where $w$ is the initial width of the current sheet, and $B_0$ is the magnetic field strength. The magnetic field is parallel to the $x$-axis, resulting in a polarity reversal around $y=0$. By counteracting the Lorentz force term with a thermal pressure gradient, we can obtain an initial equilibrium condition 
\begin{equation} \label{9}
	p = \frac{(\beta +1)}{2}B_0^2-\frac{\bm B^2}{2},
\end{equation}
where the plasma parameter is set as $\beta = 0.01$. As a result, the total pressure remains invariable throughout the current sheet. 
	
The evolution of CR particles can be described by the following equations.
\begin{equation}    \label {particle-eq1} 
	\frac{{\rm d}\bm{x}_{\rm p}}{{\rm d}t} = \bm{v}_{\rm p},
\end{equation}
\begin{equation}  \label{particle-eq2}
	\frac{{\rm d}(\gamma \bm v)_{\rm p}}{{\rm d} t} = \alpha_{\rm p}(c\bm E + {\bm v}_{\rm p}\times \bm B),
\end{equation}
where $\gamma = 1/\sqrt{1-v_{\rm p}^2/{c^2}}$ is the Lorentz factor, and $\bm x_{\rm p}$ and $\bm v_{\rm p}$ indicate the spatial position and velocity of a charged particle (proton for our scenario), respectively. When numerically solving Eqs. (\ref{particle-eq1}) and (\ref{particle-eq2}) by the Boris integrator,\cite{Boris1971} the electric and magnetic fields $\bm E$ and $\bm B$ are computed from the magnetized fluid using a cubic spline interpolation approach.

We use periodic boundaries in the $X$-direction and reflective conditions in the $Y$-direction. To numerically solve Eqs. (\ref{1}) to (\ref{5}), we choose the HLL Riemann solver with the Characteristic Tracing Contact (CT-Contact) electromagnetic field averaging scheme and the second-order piecewise linear reconstruction and second-order Runge-Kutta time step. We set the numerical resolution to $2048 ~\times ~1024$ throughout this paper, leading to a grid size of $\delta L \simeq 5{V_{\rm A}/\omega_{\rm p}}$. We uniformly inject 1 particle into each cell and initialize them with a Maxwellian distribution of the thermal velocity of $0.1V_{\rm A}$ in each direction. Based on the particle mass $m_{\rm p}$, the particle quantity can be converted into grid quantity by the ratio of particle mass to gas mass in the grid. Specifically, particles are regarded as macroparticles with a mass density of $\rho_{\rm p}$, with a default value of 1 and the same unit as the gas density $\rho_{\rm g}$. The particle mass is $\rho_{\rm p}V$, where $V$ is the grid volume. For example, in our simulation, $V = \delta L_{\rm x}\times \delta L_{\rm y}\times \delta L_{\rm z}$, where $\delta L_{\rm z} = 1$ for 2D simulation, therefore, $V = \delta L^2 = (10000/2048)^2$, then we get particle mass $\sim 0.04$. In our simulation, $\rho_{\rm plasma} = 1$, therefore, $\rho_{\rm CR}/\rho_{\rm plasma} \sim 4\%$. 

In our simulations, we set some fixed parameters, such as the uniform background plasma density of $\rho=1.0$, and Alfv\'en velocity of $V_{\rm A}=1.0$. Due to the inherent numerical dissipation of the code, we did not set explicit resistivity.
The variable parameters are listed in Table \ref{table}. We inject a velocity disturbance with a maximum amplitude of $V_{\rm eps} \sim 0.01 V_{\rm A}$ near $Y=0$ to introduce magnetic reconnection. We terminate the simulation at the final integration time of $t = 6.0\times 10^5 ~ \omega_{\rm p}^{-1}$, enough to allow all models to fully evolve. 

\begin{table}
	\begin{center}
		\setlength{\tabcolsep}{5mm}
		\begin{tabular}{ccc} 
			\hline
			\text{Models} & \text{$\bm B_{\rm g}$} & \text{Feedback} \\
			\hline
			M1   & 0.0    & OFF \\
			M2   & 0.0    & ON  \\
			M3   & 0.1    & OFF \\
			M4   & 0.1    & ON  \\
			\hline
	\end{tabular}
	\caption{Variable parameters used in our simulations. $\bm B_{\rm g}$ is the initial guide field, $Feedback$ the feedback effect of particles on the fluid.}
	\label{table}
	\end{center}
\end{table}
\section{Results}
\label{sect:Res}
\begin{figure*}
    \centering
    \includegraphics[width=0.9\linewidth]{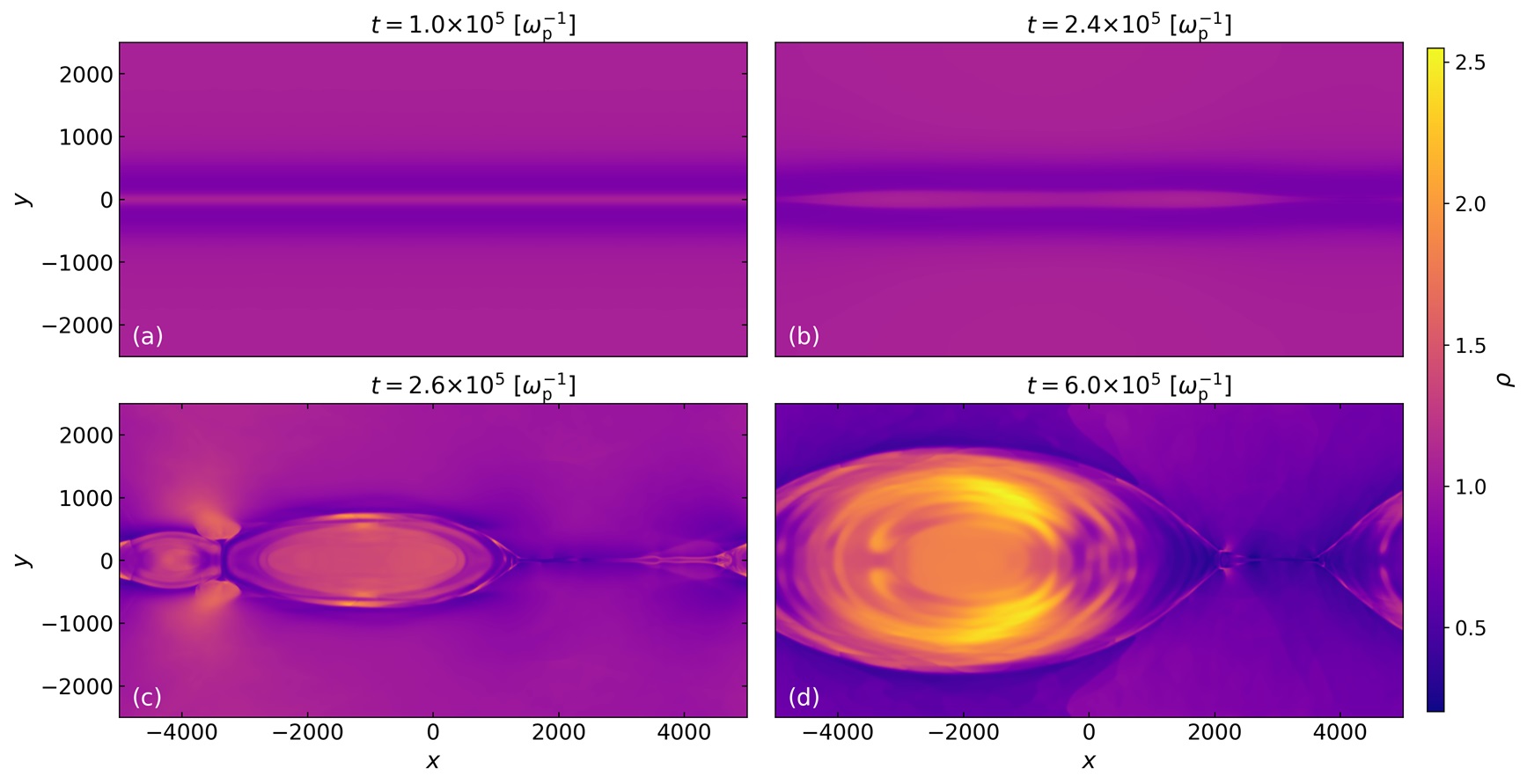}
    \caption{Plasma density (color bar) snapshots at four simulation times: (a) $t = 1.0 \times 10^5 \omega_{\rm p}^{-1}$; (b) $t = 2.4\times10^5 \omega_{\rm p}^{-1}$; (c) $t = 2.6\times10^5 \omega_{\rm p}^{-1}$ and (d) $t = 6.0\times10^5 \omega_{\rm p}^{-1}$. The data are based on the M1 model.}
    \label{fig:rho_map}
\end{figure*}
To characterize the plasma evolution during reconnection, we first analyze the temporal variation of the plasma structure, as shown in Fig. \ref{fig:rho_map}. At $t = 1.0 \times 10^5 \omega_{\rm p}^{-1}$, the current sheet structure remains largely unchanged. Subsequently, at $t = 2.4\times 10^5 \omega_{\rm p}^{-1}$, the current sheet begins to thin locally near $Y=0$, particularly at $3000.0 \leq X \leq 5000.0$ (Fig. \ref{fig:rho_map}(b)). After that, the current sheet gradually fragments, leading to the formation of multiple independent magnetic islands ($t = 2.6 \times 10^5\omega_{\rm p}^{-1}$ in Fig. \ref{fig:rho_map}(c)). These magnetic islands interact and merge over time, eventually coalescing into a single dominant magnetic island ($t = 6.0 \times 10^5\omega_{\rm p}^{-1}$ in Fig. \ref{fig:rho_map}(d)). The scale of the magnetic island at the final snapshot is slightly smaller than the size of the system, marking the saturation of the system. This evolution represents a classic 2D plasmoid reconnection process. 

\begin{figure}
    \centering
    \includegraphics[width=0.95\linewidth]{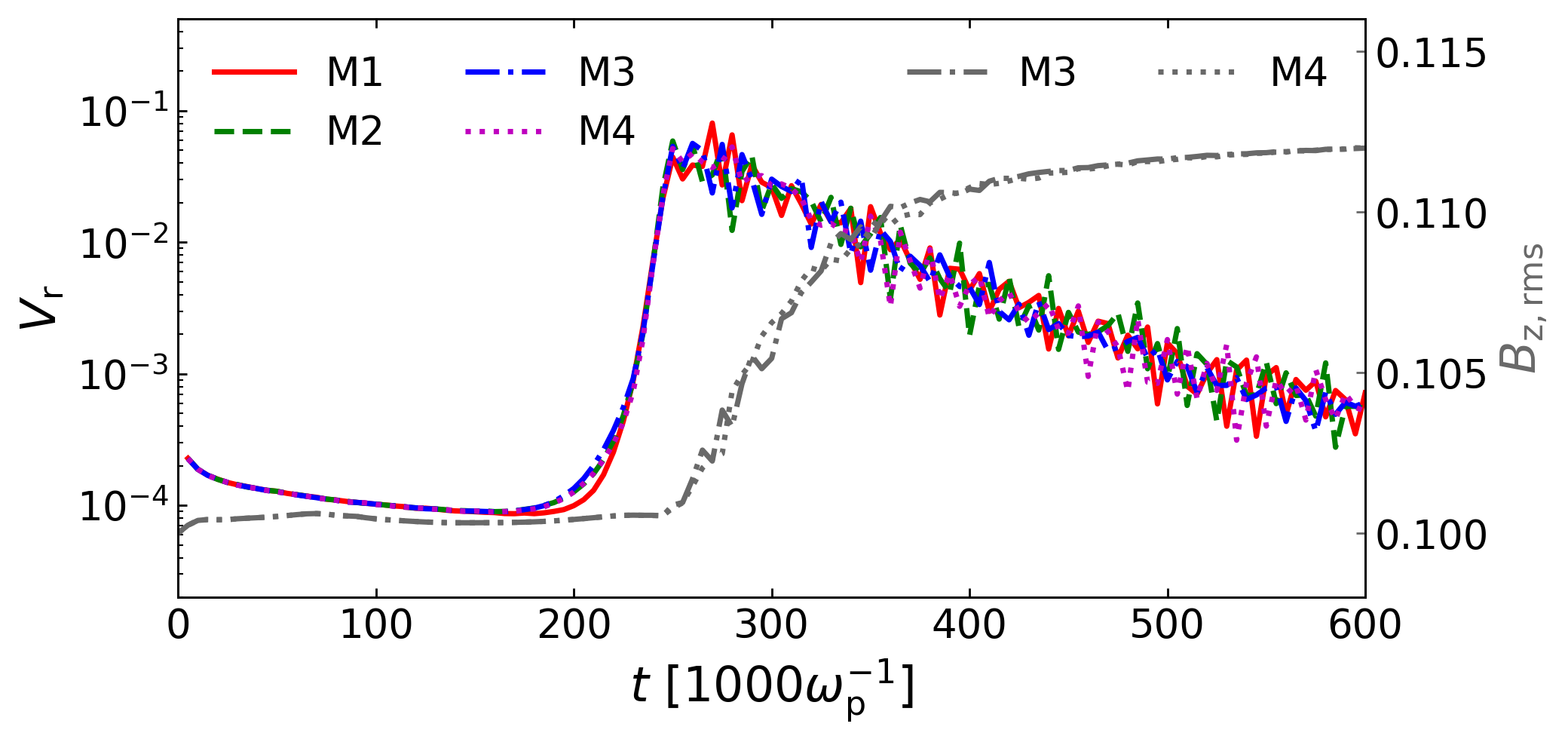}
    \caption{Reconnection rate (left axis) and $B_{\rm z,rms}$ (right axis) as functions of time.}
    \label{fig:Vrec}
\end{figure}
In the magnetic reconnection scenario, the reconnection rate is an effective parameter that characterizes the reconnection speed, defined as $V_{\rm r} = \frac{1}{B_{\rm 0}L_x}\frac{\partial}{\partial t} \int \int |\bm B_{\rm x}|dydx$.\cite{Vicentin2025}
Figure \ref{fig:Vrec} illustrates the temporal evolution of reconnection rate (left axis) for all models. 
As shown in Fig. \ref{fig:Vrec}, the reconnection rates of all four models remain nearly identical throughout the evolution. The evolution process can be divided into three distinct stages. Initially, for $t \leq 2.0\times 10^5\omega_{\rm p}^{-1}$, the reconnection rate remains low and relatively constant at approximately $10^{-4}$, indicating that the system is in a static reconnection phase where the reconnection rate is proportional to the numerical resistivity (on the order of 0.0001). This period coincides with the development of instabilities. Subsequently, during $2.0\times 10^5\omega_{\rm p}^{-1}\leq t\leq 2.5\times10^5\omega_{\rm p}^{-1}$, the reconnection rate increases rapidly, reaching a maximum value of approximately 0.1. In this case, tearing instability dominates the reconnection process, causing the current sheet to deform and fragment into magnetic islands. As these magnetic islands gradually merge, the reconnection rate peaks until a single dominant magnetic island forms ($t\geq 2.5\times 10^5\omega_{\rm p}^{-1}$). Following this peak, the reconnection rate gradually decreases as $B_{\rm x}$ decreases due to ongoing reconnection. The nearly identical evolution of reconnection rates across all four models suggests that neither the guide field nor particle feedback significantly influences the reconnection rate during system evolution.

The evolution of the root mean square of the out-of-plane magnetic field ($B_{\rm z,rms}$) further elucidates the dynamics of reconnection. Although the spatial average of the out-of-plane magnetic field component $B_{\rm z}$ of all models remains constant at the initial value due to the conservation of magnetic flux, the $B_{\rm z,rms}$ of M3 and M4 models significantly increase. This growth begins at $t \simeq 2.4 \times 10^5\omega_{\rm p}^{-1}$, occurring simultaneously with peak reconnection rate and plasmoid formation. The $B_{\rm z,rms}$ eventually saturates at a value of 0.112, indicating that its amplification is sustained by energy conversion processes during magnetic reconnection, likely through plasma compression.

\begin{figure}
    \centering
    \includegraphics[width=0.95\linewidth]{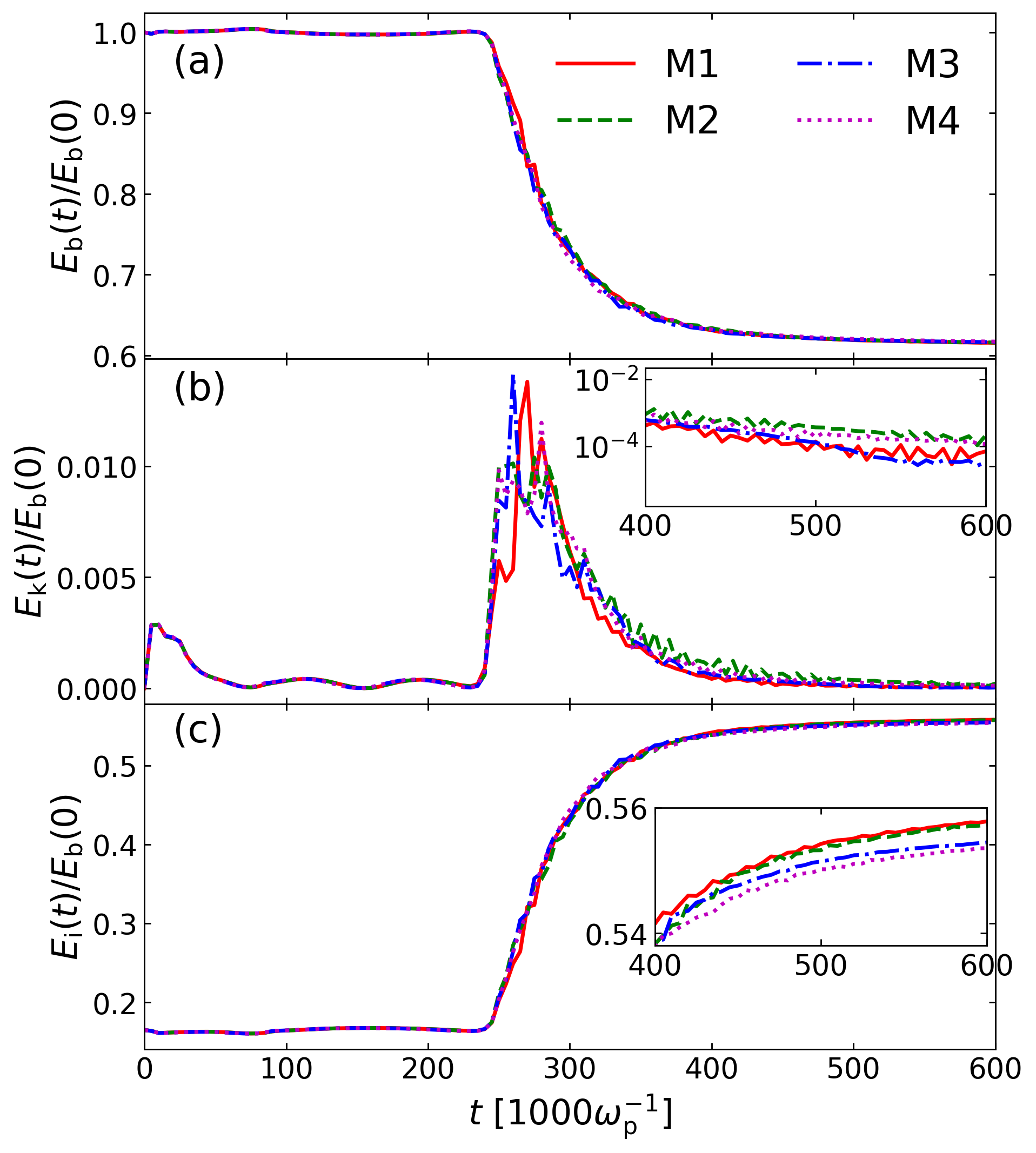}
    \caption{Magnetic energy (panel (a)), kinetic energy (panel (b)) and internal energy (panel (c)) as functions of time. The insert plots in panels (b) and (c) are enlarged versions of the curves after $t\geq 4.0\times10^5\omega_{\rm p}^{-1}$. For convenience, the total magnetic energy is calculated as the sum of the magnetic field energies of the x and y components, i.e., $E_{\rm b} = E_{\rm bx}+E_{\rm by}$.}
    \label{fig:E_gas}
\end{figure}

To further verify the source of energy growth in the guiding field, we analyze the temporal evolution of global magnetic energy (panel (a)), fluid kinetic energy (panel (b)) and internal energy (panel (c)), as shown in Fig. \ref{fig:E_gas}. Similarly to the reconnection rate, the magnetic energy evolution (panel (a)) can be divided into three stages, though its variations slightly lag behind those of the reconnection rate. The peak in reconnection rate corresponds to a phase of rapid magnetic energy dissipation, consistent with the definition of reconnection rate. The temporal evolution of kinetic energy (panel (b)) can also be segmented into three stages. After the peak value, the kinetic energy gradually decreases as a result of both the declining reconnection rate and its dissipation into internal energy. The inset in panel (b) shows that the kinetic energy is about $10^{-4}$, still significantly higher than the kinetic energy of initial velocity perturbation ($E_{\rm k}(0)\sim 1.27\times 10^{-6}$). For $t\geq 4.0\times 10^5\omega_{\rm p}^{-1}$, the kinetic energy in M2 slightly exceeds that in M1, indicating the influence of particle feedback on the fluid. The evolution trend of internal energy (panel (c)) is opposite to that of magnetic energy, its rapid increase corresponds to the rapid decrease in magnetic energy. The inset in Fig. \ref{fig:E_gas}(c) reveals that at the final time, the internal energy in M3 is lower than in M1 by approximately 0.0017, which is comparable to the energy required for guide field growth ($\sim0.0013$). This suggests that the increase in guide field energy suppresses the conversion of magnetic energy into internal energy, indicating that the guide field absorbs a portion of the magnetic energy released during reconnection, thereby limiting the growth of internal energy.

\begin{figure}
    \centering
    \includegraphics[width=1.0\linewidth]{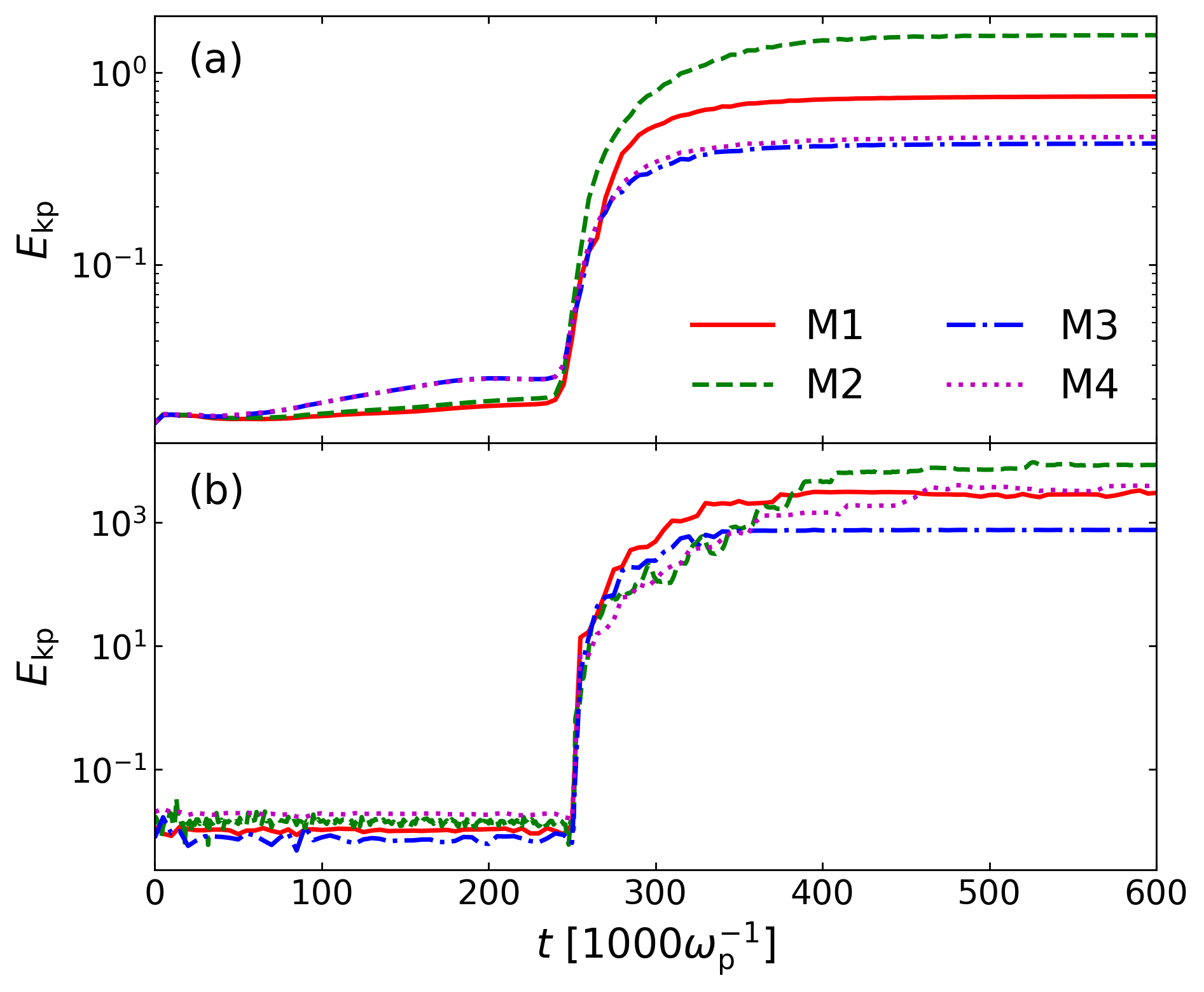}
    \caption{Kinetic energy of all particles (panel (a), showing average values) and a selected particle (panel (b)) as functions of time for four models. We select the particle with the highest energy in the final snapshot for all models.}
    \label{fig:Ek_p}
\end{figure}
To study the process of particle acceleration, we calculated the energy changes of the particles. Figure \ref{fig:Ek_p} shows the temporal evolution of the average kinetic energy for all particles (panel (a)) and a selected particle (panel (b)) for all models. The variation of all particle energies over time can be divided into three different stages:
(a) $t \leq 2.4\times 10^5~\omega_{\rm p}^{-1}$, the particle energy exhibits a slow increase.
(b) $2.4 \times 10^5 ~\omega_{\rm p}^{-1}\leq t < 2.6 \times 10^5~\omega_{\rm p}^{-1}$, the particle energy undergoes a rapid increase.
(c) $t \geq 2.6 \times 10^5~\omega_{\rm p}^{-1}$, the particle energy stabilizes, reaching a saturated state.

In the saturation stage ($t = 6.0 \times 10^5 ~\omega_{\rm p}^{-1}$), the average kinetic energy of the particles in the M2 model is higher than that of the M1 model. This significant difference highlights the role of particle feedback in improving the overall efficiency of particle acceleration. For the M3 and M4 models, the particle energy is lower than that of M1 ($\bm B_{\rm g} = 0$). This indicates that the guide field during the plasmoid reconnection stage will suppress particle acceleration. This may be due to the absorption of the reconnection field by the guide field during the plasmoid reconnection stage (as seen in Fig. \ref{fig:Vrec}). The acceleration process of individual particles in each model also exhibits a similar behavior (Fig. \ref{fig:Ek_p}(a)), but there is a period of gradual growth after rapid growth and eventually reaches saturation.

\begin{figure*}
    \centering
    \includegraphics[width=0.88\linewidth]{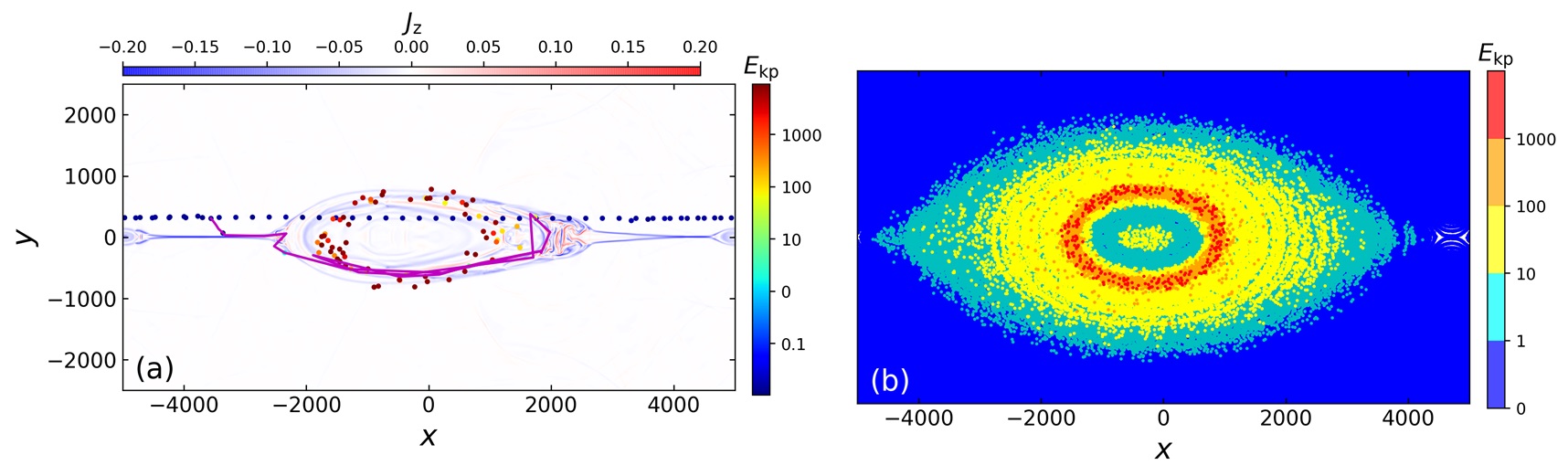}
    \includegraphics[width=0.9\linewidth]{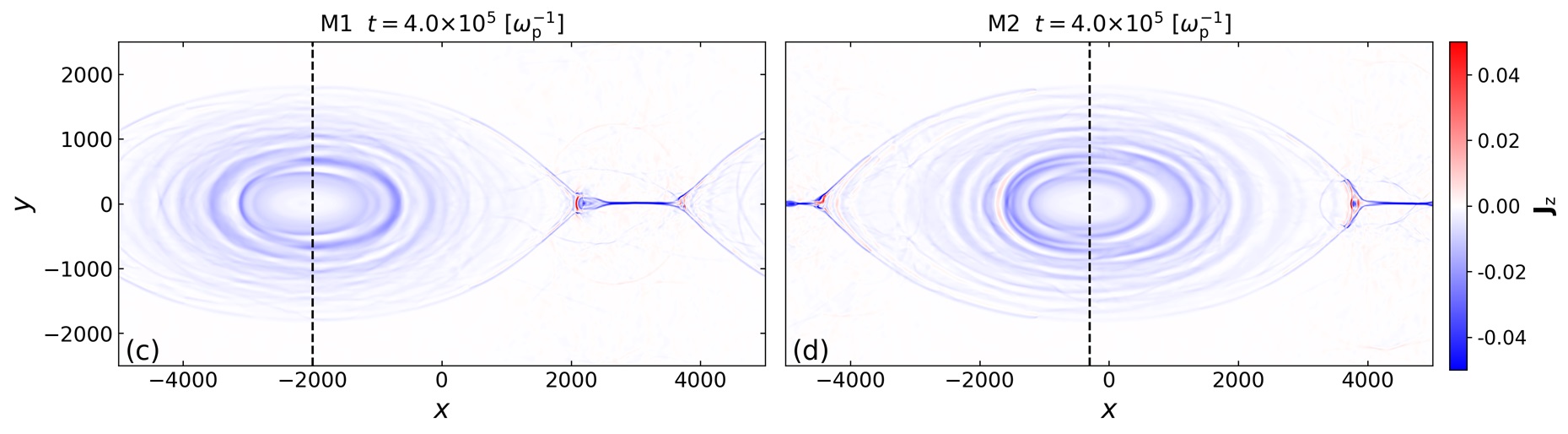}
    \caption{Panel (a): Temporal evolution of the position of a representative accelerating CR proton, with color encoding kinetic energy. These data points are sampled at uniform time intervals ($\Delta t = 5\times 10^3\omega_{\rm p}^{-1}$) cover the entire evolution period of the system. The background shows the current density distribution at $t=2.6 \times 10^5 \omega_{\rm p}^{-1}$, and the purple line denotes the particle trajectory during the peak acceleration phase ($2.5 \times 10^5 \omega_{\rm p}^{-1}\leq t \leq 2.7 \times 10^5 \omega_{\rm p}^{-1}$).
    Panel (b): The spatial distribution of particles with different energies at the final snapshot. The data are based on M2.
    Panels (c) and (d): Current density and the position distribution of the ten highest energy particles at $t = 4.0\times10^5~\omega_{\rm p}^{-1}$. The dashed line represents the centerline of the magnetic island in the y-direction. The data are based on M1 and M2 models.}
    \label{fig:E&particles}
\end{figure*}

Figure \ref{fig:E&particles}(a) shows the position of an accelerating particle changes with time, and the color represents the particle energy. The background color is the current density distribution when $t = 2.6\times 10^5\omega_{\rm p}^{-1}$, and the purple line is the particle trajectory during the period of $2.5\times 10^5\omega_{\rm p}^{-1}\leq t \leq 2.7\times 10^5\omega_{\rm p}^{-1}$. In the early phase of the simulation, the particle exhibits quiescent motion with minimal displacement in the y-direction and no significant energy gain. At $t = 2.5 \times 10^5 \omega_{\rm p}^{-1}$, the particle enters the current sheet, where it begins to gain energy rapidly. After that, it is advected into the magnetic island and becomes trapped. Subsequently, the particle circulates within the island to its right side, where its path is deflected, likely by magnetic structures during a merger event. Following this deflection, the particle remains confined within the magnetic island, continuing its circulation. This visualization directly demonstrates the process of particle trapping and acceleration within magnetic islands.
The spatial distribution of all particles at the final simulation time ($t = 6.0 \times 10^5\omega_{\rm p}^{-1}$) is presented in Fig. \ref{fig:E&particles}(b), revealing a highly structured energy-dependent configuration. The highest-energy particles (red) are predominantly confined within a ring-like structure inside the magnetic island, with a semi-minor axis in the $y$-direction of approximately $800V_{\rm A}/\omega_{\rm p}$, which is consistent with the maximum $y$-displacement observed in Fig. \ref{fig:E&particles}(a). Medium-energy particles (yellow and green) populate the central region of the island and form a transitional shell outside the high-energy ring. In contrast, the majority of low-energy particles (light blue) are dispersed widely in the periphery of the simulation domain.

\begin{figure}
    \centering
    \includegraphics[width=0.95\linewidth]{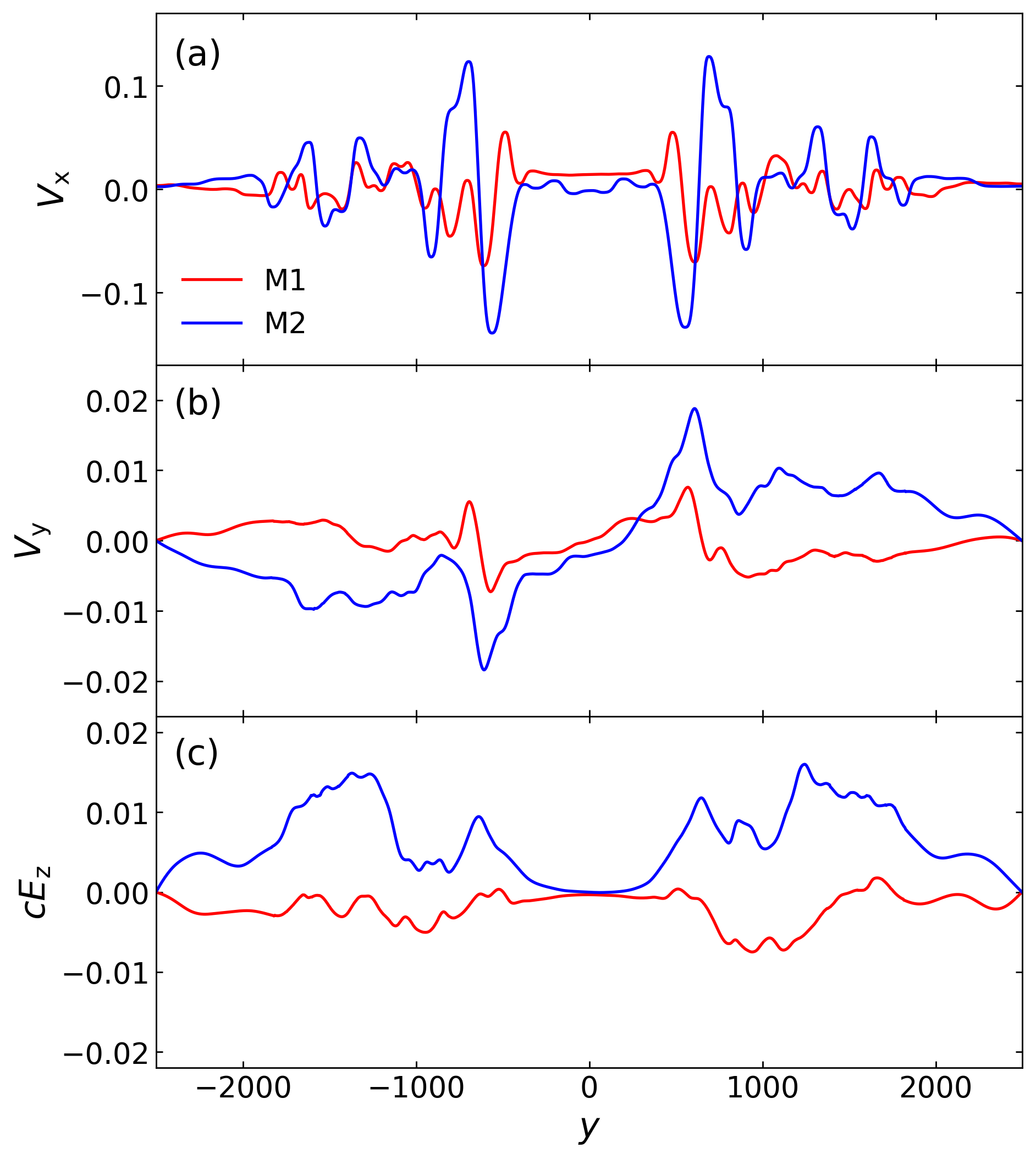}
    \caption{Variations of the fluid velocity ($\bm V_x$), panel (a) and panel (b) for $\bm V_y$) and electric field ($c\bm E_z$, panel (c)) along the y-axis at the center of the magnetic island, corresponding to the dashed lines in Figs. 5(c) and (d).}
    \label{fig:Vx_y}
\end{figure}
At $t = 4.0\times 10^5~\omega_{\rm p}^{-1}$, both models show their highest-energy particles confined within the magnetized islands. At this stage, the acceleration of the highest energy particle in M1 has ended, while the highest-energy particle in M2 continues to gain energy (Fig. \ref{fig:Ek_p}(b)). This means that the particle acceleration efficiency of the two models in the magnetic island is different; to explain the above differences, Fig. \ref{fig:Vx_y} presents the variation of the fluid velocity ($V_x$ and $V_y$ correspond to panels (a) and (b)) and electric field $cE_{\rm z}$ (panel (c)) at the center of the magnetic island along the y-axis direction (corresponding to the dashed lines in the lower row of Fig. \ref{fig:E&particles}). At the center of the magnetic island, $V_x \sim 0, V_y \sim 0$, and as it moves away from the center of the magnetic island, the amplitude of velocity rapidly increases and decreases, indicating strong shear flow in the magnetic island. Here, the “shear flow” is defined as the uneven, circulatory motion that develops around the magnetic island center. This flow exhibits a layered structure (as shown in Figs. 5(c) and (d)), in which adjacent layers move at distinct velocities and generate a persistent velocity shear. Quantitative analysis shows that the sum of shear velocity amplitudes ($V=\sqrt{\langle V_x^2+V_y^2\rangle}$ is calculated using the data in Figs. \ref{fig:Vx_y}(a) and (b)) in the M2 model is approximately 2.16 times that of the M1 model, and for the electric field, this ratio is approximately 3.14. This confirms that the feedback-enhanced shear flow makes a significant contribution to the total electric field responsible for particle acceleration.
This indicates that particle feedback in M2 improves the shear flow within the magnetic island, and the shear flow inside the magnetic island can effectively enhance the convective electric field $\bm cE = -\bm v_{\rm g} \times \bm B$, thus allowing particles to reach higher energy within the magnetic island.

\begin{figure*}
    \centering
    \includegraphics[width=0.9\linewidth]{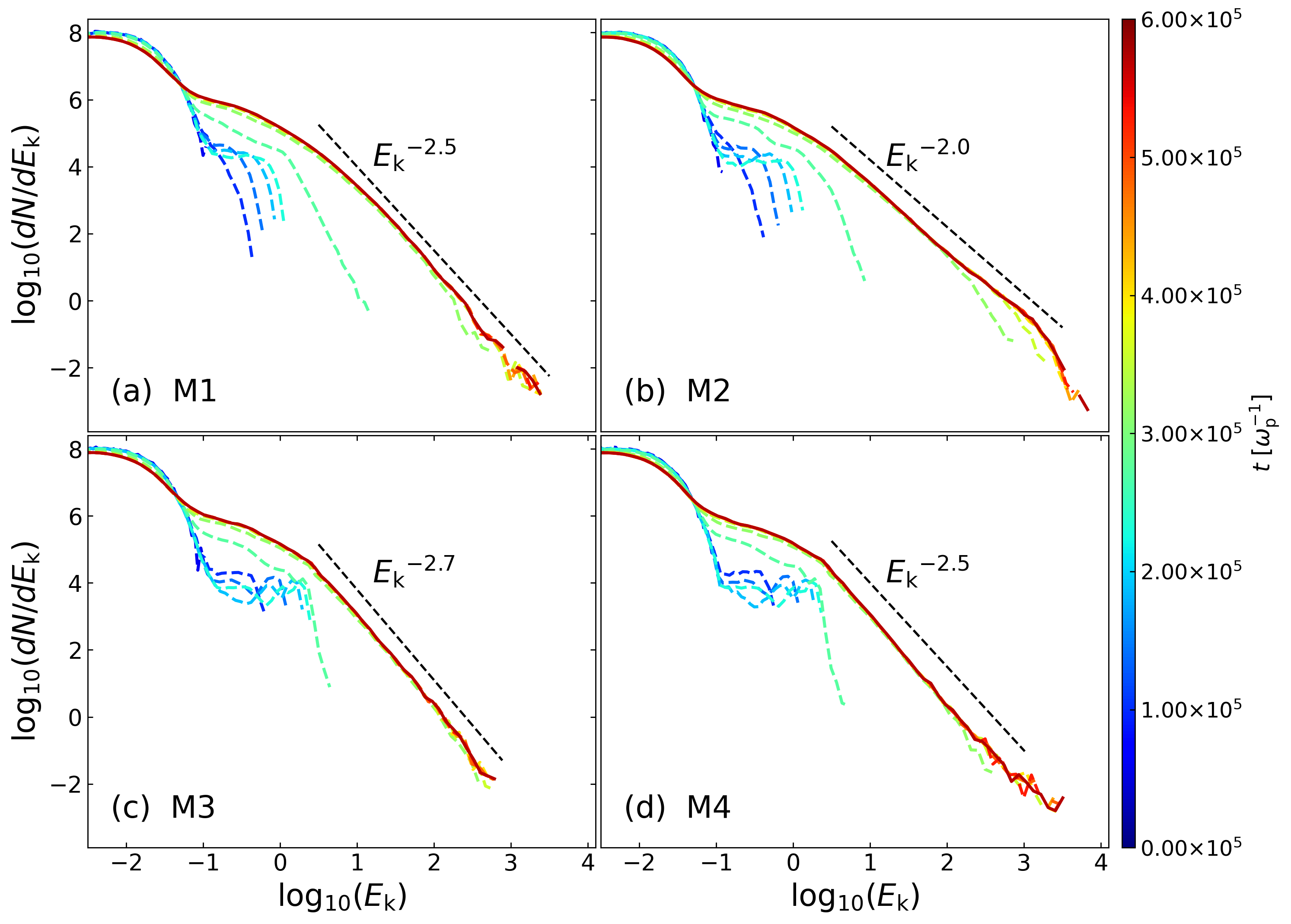}
    \caption{The energy spectra of all particles at different evolutionary times (color bar) for all models. The particle distributions at the final snapshots are marked with solid lines.}
    \label{fig:Ek_sepctra}
\end{figure*}
Figure \ref{fig:Ek_sepctra} presents the particle energy spectra for the four models with color indicating the progression of time. The particle distributions in all four models converge to a power-law form, demonstrating the effectiveness of reconnection acceleration. The energy spectrum of M2 is slightly shallower than that of M1. This is attributed to the modulation of the convective electric field by particle feedback, which enhances the efficiency of particle acceleration. Furthermore, the maximum particle energy in M2 is approximately half an order of magnitude higher than in M1, consistent with the results of Fig. \ref{fig:Ek_p}.

The energy spectrum of M3 is steeper than that of M1, and the maximum particle energy in M3 is lower than that in M1. This indicates that the guide field suppresses particle acceleration, which is consistent with Fig. \ref{fig:Ek_p}. Similar to the zero-guide field cases, the spectral exponent of M4 is smaller than that of M3, and the maximum particle energy in M4 is higher than that in M3, reaching a level similar to that of M1. This indicates that particle feedback can still effectively promote particle acceleration in the presence of a guide field.

\section{Discussion}\label{Sec:Discussion}
Current numerical studies on magnetic reconnection adopt primarily three approaches. The first employs MHD+test particle approximations to simulate magnetic reconnection on large scales, which neglect the kinetic behavior of individual particles.\cite{Gordovskyy2010a,Gordovskyy2010b,Kowal2011} The second utilizes first-principles-based PIC simulations to study electron and ion dynamics at kinetic scales, providing detailed insight into particle behavior but struggling to capture macroscopic plasma dynamics.\cite{Sironi2014,Guo2014,Guo2015} 
The third approach employs hybrid simulations that bridge the gap between MHD and fully kinetic methods, capturing ion kinetic effects while maintaining computational efficiency for larger systems.\cite{Walia2022JGRA,Shi2023JGRA,Jain2024AA,Zhang2024PRL} In this study, we carried out the numerical simulations of magnetic reconnection and explored the role of particle feedback in the reconnection acceleration on the transition scale between the large MHD scale and the small kinetic plasma one, by using the MHD-PIC method.

Our simulations begin with a Harris current sheet configuration, and we add noise to velocity fluctuations to trigger magnetic reconnection with an initial magnetic field reversal. Unlike the MHD plus test-particle method,\cite{Liu2009,Kowal2011,Gordovskyy2010a,Ripperda2017a} where particles evolve on a frozen fluid background, our model features coevolution of particles and the fluid, fully accounting for particle feedback effects. This allows for real-time updates of electromagnetic fields, making our simulations more realistic. Although the particle feedback model exhibits minimal differences in fluid evolution compared to the non-feedback model, the primary effect of particle feedback is to adjust the flow field details to enhance the shear flow inside the magnetic islands, thereby enhancing the convective electric field $\bm cE = -\bm v_{\rm g} \times \bm B$ inside the magnetic islands. 

\citeauthor{Seo2024} investigated particle feedback within a co-evolving particle-fluid framework using the Fokker-Planck (FP) equation,\cite{Seo2024} considering the collective feedback of all particles. Their results indicate that a higher ratio of nonthermal particles to fluid density leads to a steeper particle energy spectrum. We conducted two additional models by adjusting the number of particles injected into each grid cell to change $R$, and obtained similar results (not shown here). In their model, the distribution of nonthermal particles strongly depends on the diffusion model, and the feedback effect of particles is limited by the particle distribution and diffusion coefficient in the FP equation. When considering an isotropic particle diffusion model, the density distribution structure of particles will not exist. While in the MHD-PIC model, the particle distribution does not depend on the particle diffusion model. Although there are differences between the two models, similar results were obtained, indicating the rationality of the MHD-PIC model in handling particle feedback.

\section{summary} \label{Sec:Summary}
By using the MHD-PIC method, we performed numerical simulation of reconnection acceleration of particles. In the framework of the coevolution of fluid and particles, we focus on exploring the impact of particle feedback on particle acceleration in magnetic reconnection processes. Our numerical findings are summarized as follows:
\begin{enumerate}

 \item The particle energy spectrum eventually converges to a power-law distribution, validating the effectiveness of reconnection acceleration.
	
 \item Particle feedback improves the efficiency of particle acceleration by enhancing the shear flow inside the magnetic island. This results in a hardening of the particle energy spectrum, with the maximum particle energy in feedback models exceeding that of non-feedback models.
	
 \item The guide field suppresses both the increase in gas internal energy and particle acceleration. 
\end{enumerate}	

\begin{acknowledgements}
 We thank the anonymous referee for valuable comments that significantly improved the quality of the paper. This work is supported in part by the High-Performance Computing Platform of Xiangtan University. N.Y. Y. is grateful for the support from the National Nature Science Foundation of China (No. 12431014) and the Project of Scientific Research Fund of the Hunan Provincial Science and Technology Department (2024ZL5017). S. M. L. is grateful for the support from the Xiangtan University Innovation Foundation for Postgraduate (No. XDCX2024Y164).
\end{acknowledgements}

\nocite{*}
\bibliography{aipsamp}

\end{document}